\documentclass[aps,prl,twocolumn,superscriptaddress,nofootinbib,10pt]{revtex4-2}

\usepackage{graphicx} 
\usepackage{dcolumn} 
\usepackage{bm} 
\usepackage{slashed}
\usepackage{xcolor}
\usepackage{xspace}
\usepackage{amsmath}
\usepackage{amsfonts}
\usepackage{lmodern}
\usepackage{tabulary}
\usepackage{dsfont}
\usepackage{hyperref}
\usepackage{placeins}
\usepackage[letterpaper,top=1.0in, inner=0.5in, outer=0.5in,bottom=1.0in,,headheight=0pt,headsep=20pt,centering]{geometry}

\newcommand{\model}{\textsc{Passwd-ABC}\xspace}
\newcommand{\abc}{\textsc{ABC}\xspace}
\newcommand{\loss}{\mathcal{L}}
 \graphicspath{{figures/}}
 
\begin{document}

\title{A data-driven and model-agnostic approach to solving combinatorial assignment problems in searches for new physics}

\author{Anthony Badea} 
\email{anthony.badea@cern.ch}
\affiliation{Harvard University, Cambridge, Massachusetts, 02138}%
\affiliation{University of Chicago, Chicago, Illinois, 60637}%

\author{Javier Montejo Berlingen} 
\email{jmontejo@cern.ch}
\affiliation{Instituto de Física de Altas Energías, Campus UAB, 08193, Bellaterra (Barcelona), Spain}
\affiliation{QUP, KEK, 1-1 $\bar{O}$ho, Tsukuba, Ibaraki 300-3256}
\date{\today}

\begin{abstract}
We present a novel approach to solving combinatorial assignment problems in particle physics without the need to introduce prior knowledge or assumptions about the particles' decay.
The correct assignment of decay products to parent particles is achieved in a model-agnostic fashion by introducing a novel neural network architecture, \model, which combines a custom layer based on attention mechanisms and dual autoencoders.
We demonstrate how the network, trained purely on background events in an unsupervised setting, is capable of reconstructing correctly hypothetical new particles regardless of their mass, decay multiplicity and substructure, and produces simultaneously an anomaly score that can be used to efficiently suppress the background.
This model allows to extend the suite of searches for localized excesses to include non-resonant particle pair production where the reconstruction of the two resonant masses is thwarted by combinatorics.
\end{abstract}
\maketitle


Searches for beyond the Standard Model (BSM) physics at the LHC are producing increasingly stringent bounds on the most theoretically appealing models \cite{ATLAS:2023tkt,ATLAS:2018sbw,ATLAS:2019wdu,CMS:2019kaf,CMS:2021eha,ATLASpublic,CMSpublic}. This has led to a paradigm shift in the program towards more signature-based rather than model-based searches. The archetype of this approach is the search for an excess in the di-object mass spectrum, which has provided a fruitful history of discoveries, from the $J/\psi$ meson~\cite{E598:1974sol,SLAC-SP-017:1974ind} to the Higgs boson~\cite{ATLAS:2012yve,CMS:2012qbp}. 
%
An important shortcoming of this approach is that it is restricted to models with resonant production of a single particle, from which all the relevant final-state objects are assumed to originate.
However, many BSM models lead to non-resonant pair production which faces the additional challenge of assigning the decay products to parent particles in order to reconstruct masses.
This challenge is further exacerbated in decays to high-multiplicity final states. Heuristic approaches have been applied in final states with low object multiplicities, such as di-jet pairs or di-$b$-jet pairs~\cite{ATLAS:2017jnp,ATLAS:2018tti,ATLAS:2023qzf,ATLAS:2023ssk}, where all possible combinations can be tested. Some common choices are minimizing the sum of distances or the mass asymmetry. These approaches do not scale easily to higher object multiplicity as the amount of combinations grows in a factorial way, leading to a worsening in the resolution of the reconstructed mass together with a rapid increase in complexity beyond CPU capacity. Though machine-learning approaches have been proposed to tackle this problem, they always start with a fixed signal model assumption~\cite{Fenton:2020woz,Qiu:2022xvr,Badea:2022dzb,Ehrke:2023cpn,Shmakov:2021qdz,Lee:2020qil}. This restricts the model to learn the specific mass, multiplicity and structure of the decay. Furthermore, by imposing the signal hypothesis in the final state topology, the background spectrum is sculpted to become strongly signal-like. 

In this letter, we present a methodology to reconstruct the four-momenta of pair-produced particles by assigning decay products to parent particles without knowledge of the decay mode. This is accomplished by training a custom neural network (NN) directly on data in an unsupervised approach, outputting an object-to-parent assignment and an anomaly score. The only assumption embedded in the model is pair-production of new particles with identical decays. Beyond this, the significant computational innovations are:
\begin{itemize}
    \item A permutation equivariant and easily scalable NN layer tailored to address combinatorial assignment problems based on the attention mechanism~\cite{vaswani2023attention}, dubbed \emph{Attention-Based Combinatorial} layer, or \abc layer. The \abc layer provides a per-object probability to be assigned to a certain category.
    \item A model based on \abc layers, named \emph{Particle ASSsignment for unknoWn Decays}, \model. It is trained in an unsupervised setting directly on data events, without any signal hypothesis or input from simulation, and is capable of assigning accurately the decay products of pair-produced particles. The performance of the model is shown to be insensitive to the BSM particle masses and decay structure, and capable of reconstructing high-multiplicity final states.
\end{itemize}

Formally, given an input set $\mathcal{X} = \{x_1, \dots, x_N\}$ and a category set $\mathcal{Y} = \{y_1, \dots, y_C\}$, the goal is to assign to each input $x\in\mathcal{X}$ a label $y\in\mathcal{Y}$, $l(x_i)=y_j$. 
This defines $C$ exclusive subsets $P_k = \{x\in\mathcal{X}: l(x)=y_k, \forall x \in \mathcal{X}\}$. 
A combinatorial assignment is a function $f: \mathcal{X} \rightarrow A = \{0,1\}^{N\times C}$, and for the correct assignment $a_{ij} = \mathbb{I}[l(x_i)=y_j]$, where $\mathbb{I}[\cdot]$ is an indicator function that is one when $\cdot$ is true and zero otherwise.
Our goal is then to build a model that provides a differentiable approximation to $f \sim f^\prime: \mathcal{X} \rightarrow A^\prime = [0,1]^{N\times C}$, which can be mapped back to the hard combinatorial assignment if desired via $a_{ij} = \text{argmax}_{j \in C}(a^\prime_{ij})$.
In our setup, each of the two parent particles is considered a category ($P_1, P_2$). An additional category can be introduced to account for objects not originating from the decay products, such as initial-state radiation (ISR), underlying event, or pileup. Setups with a higher number of categories, for example targeting four-top production, are also possible but not considered here for simplicity.
The parent particles are reconstructed as $p_j = \sum_{i}^N a_{ij}^T \cdot x_i$, which also enables a differentiable approximation to the assignment of objects to parent candidates by introducing $a^\prime$ instead. Other differentiable approximations for the object assignment such as Gumbel-Softmax~\cite{maddison2017concrete,jang2017categorical} were tested and found to yield worse performance.

If a signal model is assumed, the assignment problem can be trivially framed in a supervised learning approach. Objects can be assigned a label based on truth-level information and the loss function is simply the binary or multi-class cross-entropy loss summed over all objects. This strategy, adopted by previous works~\cite{Fenton:2020woz,Qiu:2022xvr,Badea:2022dzb,Ehrke:2023cpn,Shmakov:2021qdz,Lee:2020qil}, requires training the model on signal simulation, inducing a strong model dependence and shaping of background distributions. 

In order to adapt the problem to an unsupervised learning strategy, the problem is reframed as finding the combinatorial assignment such that the \emph{similarity} between the two particle sets is maximized, $P_1 \approx P_2$. 
The choice of similarity metric is therefore the key ingredient to this setup.
Different heuristic approaches have been used in the past in analyses where the number of combinations allows for a brute-force approach, such as the grouping with minimum mass asymmetry, $\mathcal{A}=\frac{|m_1-m_2|}{m_1+m_2}$. Such similarity metrics can also be introduced in our setup which would lead to a CPU-efficient way of approximating the result from iterating over all combinations, but the performance of the reconstructed variable would never surpass the heuristic approach. Instead, as detailed below, the chosen similarity metric is taken to be the distance between both parent particles in a learned feature space, allowing the model to identify and build better features to quantify the similarity. 

\begin{figure}[ht]
    \centering
    \includegraphics[width=0.35\textwidth]{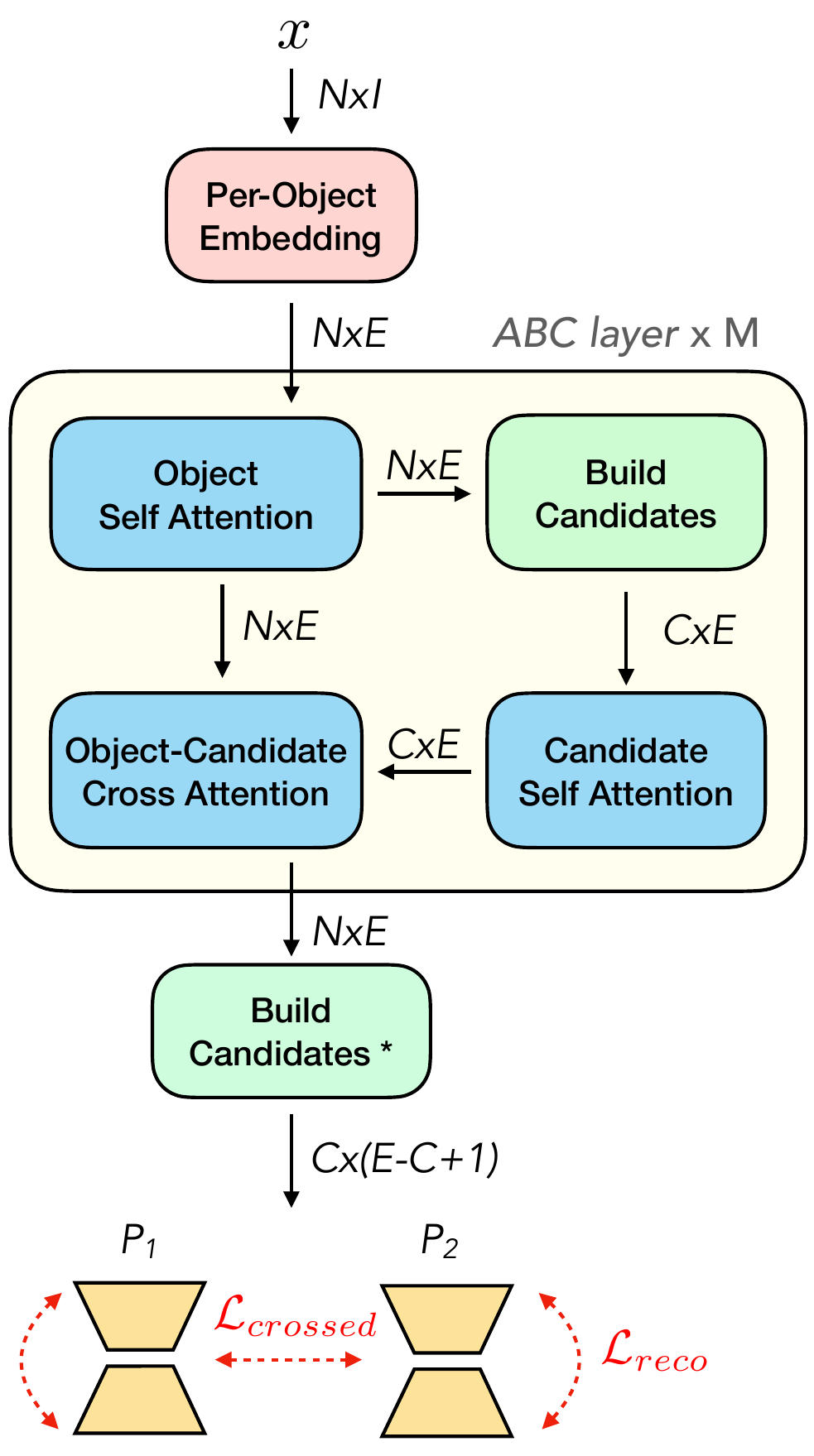}
    \caption{Sketch of the \model model and components of the \abc layer. Multiple \abc layers can be stacked to build the model. A final candidate-building operation is performed where the $C$ first features encoding the category probability are dropped and the mass of the candidate is added. This alternative candidate building block is notated with an asterisk. Red lines indicate the terms that are used to build the reconstruction loss and similarity loss.}
    \label{fig:model_diagram}
\end{figure}

The \abc layer and \model model composition are shown in Figure~\ref{fig:model_diagram}. The input to the model, $x \in \mathbb{R}^{N \times I}$, consist of $N$ objects with $I$ input features.  The input objects are fed through NNs to produce learned embeddings per object, $e = \left(\text{NN}(x_1), \dots, \text{NN}(x_N)\right) \in  \mathbb{R}^{N \times E}$, with $E$ the embedding dimension. At various points in the model, the $C$ first embedded features per object will be softmax-ed to represent the probability for each object to be assigned to a given category: $a^\prime = \sigma (e^\prime \subset e) \in \mathbb{R}^{N \times C}$, with $\sigma$ the softmax function acting over the category dimension. 

The \abc layer consists of three individual attention blocks. The first is an object self-attention block, after which category probabilities are computed and candidate parents are built in the embedded space by summing all the input objects weighted by their category probabilities, $p_j^{(e)} = \sum_{i}^N {a_{ij}^\prime}^T \cdot e_i$.
The parent candidates are input to the second self-attention block, followed by an object-candidate cross-attention block. 
This structure allows the layer to learn relations among objects, candidates, and object-candidate pairs.
The input and output of the \abc layer have the same $(N,E)$ dimension, which allows multiple such layers to be stacked. The \abc layer is permutation equivariant, therefore avoiding the need for arbitrary sorting of the input objects, and alleviating problems in supervised training applications due to varying class imbalances at each input position.

Depending on the use case, another candidate-building block can be appended as a last step, summing objects in the learned feature space or Lorentz space to obtain parent candidate four-momenta. This architecture allows great flexibility as the training target can be defined at the per-object level, $e.g.$ classification of truth-labelled objects; at the particle candidate level in Lorentz space, $e.g.$ regressing the BSM particle mass; or even mixtures of learned features and Lorentz features as described below. Note that even when a training target such as BSM particle mass is defined, the model output of the object-particle assignment is still available. 

The \abc layer is used as the foundation block for \model. The additional ingredient is an autoencoder block which is applied \emph{identically} to each particle candidate. Particle candidates are built using the same approach as in the \abc layer. In addition, the reconstructed mass is added as an extra feature of the parent particle and the category scores are dropped. The autoencoder compresses the dimensionality of the embedded feature space to a latent space, $z = \{AE^\text{enc}(p_j): j \in C\} \in \mathbb{R}^{C\times L}$, summarizing each particle candidate's properties, and then attempts to reconstruct the original inputs, $\hat{p} = \{AE^\text{dec}(z_j): j \in C\}$. The distance in latent space between the two candidates, $\lVert z_1 - z_2\rVert$, is used as the metric for similarity that is required to solve the problem. 
Other approaches such as minimizing the energy mover's distance~\cite{Komiske:2019fks,Komiske:2020qhg} between the two candidates were also examined but found to have worse performance.

More formally, the loss function defined to train \model is:
\begin{align*}
    \loss &= \loss_{reco} + \loss_{crossed} + \loss_{random} (+ \loss_{ISR})  \\
    \loss_{reco} &= \lambda_{reco} \cdot \left(\Vert p_1 - \hat{p}_1\rVert + \lVert p_2 - \hat{p}_2\rVert\right) \\
    \loss_{crossed} &= \lambda_{crossed} \cdot \lVert z_1 - z_2\rVert \\
    \loss_{rand} &= \lambda_{rand} \cdot \max\left(0, 1+\lVert z_1 - z_2\rVert-\lVert z^{rand}_1 - z^{rand}_2\rVert \right)\\
    \loss_{ISR} &= \lambda_{ISR} \cdot E_T(P_{ISR})/\text{GeV} \\
\end{align*}

\noindent Where $\loss_{reco}$ is the typical autoencoder reconstruction loss and $\loss_{crossed}$ is the distance between both reconstructed particles in the autoencoder latent space.
Since all of the candidate features are learned, a simple minimization of these two terms will lead to a collapse of zero-valued candidate input features and latent space. This is avoided by adding an additional term, $\loss_{random}$, based on the triplet loss~\cite{Schroff_2015}. Two additional parent particles are built using a random combination of objects, and the distance between them is maximized. This term is implemented in the form of a hinge loss to avoid divergences where the NN ignores the main task and focuses on maximizing the difference among random candidates. An extra term, $\loss_{ISR}$, is introduced when the additional ISR category is considered and is simply the transverse energy, $E_T$, of the ISR system in units of GeV. This term is introduced to avoid all objects being flagged as ISR which leads to perfectly identical, and empty, parent particles. All loss terms are multiplied with a corresponding hyperparameter to balance the relative weight.


In order to demonstrate the generality of the approach and to quantify the performance of the model we focus on the multi-jet final state with varying decay structures. We use a simplified $R$-parity-violating supersymmetry model~\cite{Barbier:2004ez,Mohapatra:2015fua} with a nonzero baryon-number-violating $\bar{U}\bar{D}\bar{D}$ coupling~\cite{Csaki:2011ge}. We consider gluino, squark, and stop pair production ($\tilde{g}$, $\tilde{q}$, $\tilde{t}$) with cascade decays among themselves or to the lightest neutralino ($\tilde{\chi}^0_1$). The lightest supersymmetric particle then decays through the $\bar{U}\bar{D}\bar{D}$ coupling to SM quarks, leading to final states with 6 to 12 jets and different internal substructures. An example signal is $\tilde{g}\tilde{g}\rightarrow 2 \times qq \tilde{\chi}^0_1 \rightarrow 2 \times qq (qqq)$. For the sake of generality, and to emphasize the model-agnostic focus of this letter, the following notation is used: $XX^{m_X} \rightarrow 2 \times Nj (Mj)$, to denote pair-production of particle $X$ with mass $m_X$ decaying to $N$ jets plus an intermediate particle, which in turn decays to $M$ jets, producing a final state with $2\times (N + M)$ decay products. The signals, decays, and number of final state jets considered are given in Table~\ref{tab:signals}, ranging broadly over event topologies.

\begin{table}
\begingroup
\setlength{\tabcolsep}{10pt} 
\renewcommand{\arraystretch}{1.5} 
\begin{tabular}{|rll|c|}
\hline
   \multicolumn{3}{|c|}{Decay} & Number of jets \\
   \hline
   $\tilde{g}$ &$\rightarrow q \tilde{q}        $ & $\rightarrow q (qq)$  & 6 \\
   $\tilde{q}$ &$\rightarrow q \tilde{\chi}^0_1 $ & $\rightarrow q (qqq)$  & 8 \\
   $\tilde{q}$ &$\rightarrow W/Z/H \tilde{q}'   $ & $\rightarrow qq (qq)$  & 8 \\
   $\tilde{g}$ &$\rightarrow qq \tilde{\chi}^0_1$ & $\rightarrow qq (qqq)$  & 10 \\
   $\tilde{t}$ &$\rightarrow t \tilde{\chi}^0_1 $ & $\rightarrow qqq (qqq)$  & 12 \\
\hline
\end{tabular}
\endgroup
\caption{Benchmark production modes and decays that are considered within the context of R-parity-violation SUSY. Hadronic decays of bosons and top quarks are considered as they have the highest branching fractions. No jet-flavor information is used in the model, therefore $b$-quarks are not explicitly notated.}
\label{tab:signals}
\end{table}

Simulated signal events are generated at leading-order using Madgraph \cite{Alwall:2014hca, Fuks:2012im} with up to one additional parton, and interfaced with Pythia 8 \cite{Sjostrand:2014zea}. Background QCD events are simulated using Pythia 8. Both signal and background events are interfaced with Delphes event reconstruction \cite{Alwall:2014hca, Sjostrand:2014zea}, and required to have at least 6 jets with {$p_T > 25$ GeV} and {$H_T>1000$ GeV} (scalar sum of jet $p_T$) to mimic hadronic trigger selections at the LHC experiments. A sample of about 200k background events is retained after selection, where 90\% is used for training and the other 10\% is reserved for evaluation. A sample of 10k events is generated for each signal and none of the signal events are used in the training.

A minimal \model model is implemented using a stack of two \abc layers, the first one with two categories and the second with three. This choice allows for a first splitting into two candidate particles which is then refined in the second layer by trimming some of the jets and assigning them to the ISR category. All activation functions are rectified linear units (ReLu). Embedding is a three-layer feed-forward network (FFN) with a 16-feature embedding space. All attention blocks are 4-headed attention layers, followed by a three-layer FFN, using the same 16-feature dimensionality.
The autoencoder uses four layers for each of the encoder and the decoder, mapping into a 2-dimensional latent space. All hyperparameters in the loss function are set to unity except for $\lambda_{reco}=10$.
Only the jet four-momenta are used as input features, parameterized as $x_i = (\log{p_T}, \eta, \sin{\phi}, \cos{\phi}, \log{E})$. Additional features such as particle type or identification scores can be trivially added but are not considered here. Up to 12 jets are used as input and events with fewer than 12 are zero-padded.
The model is trained for 10 epochs using the ADAM optimizer~\cite{kingma2017adam}, with a batch size of 1024. The learning rate is warmed up for the first 2\% of training steps to a maximal value of $10^{-3}$ and then decayed by 95\% every further 2\%. 

\begin{figure}[t]
\centering
\includegraphics[width=0.45\textwidth]{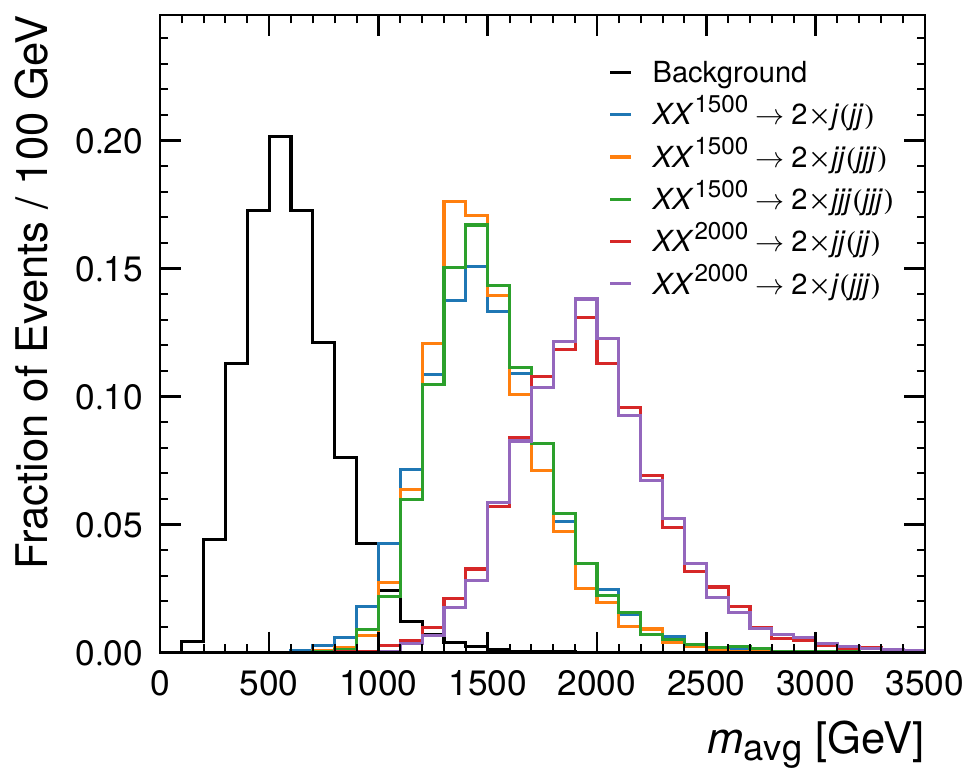} \\ 
\includegraphics[width=0.45\textwidth]{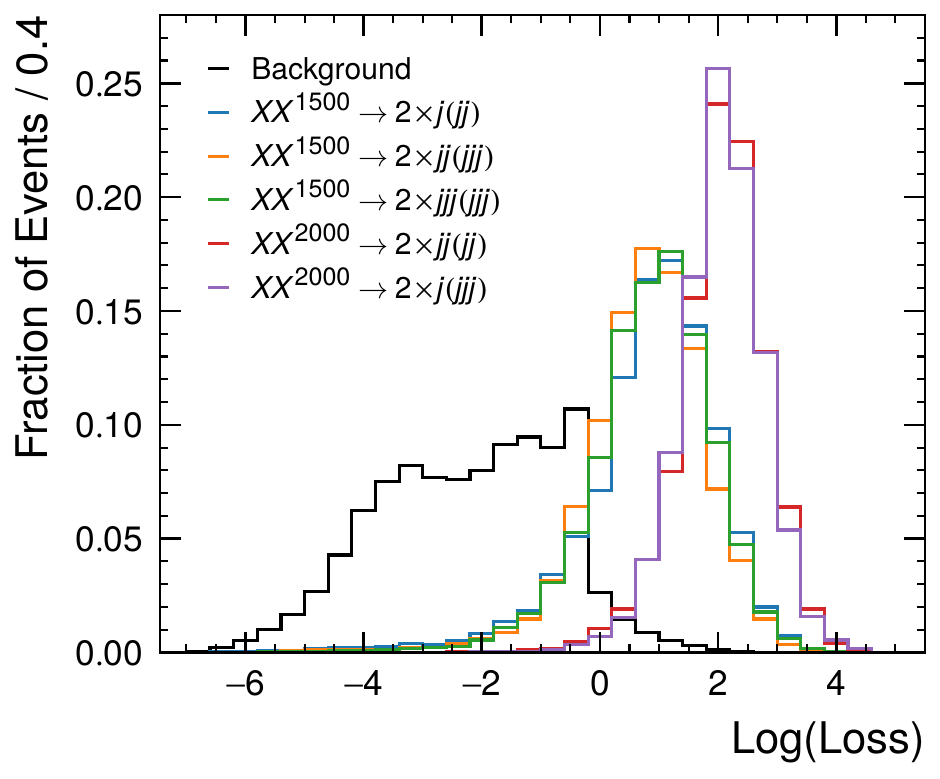} 
\caption{Average reconstructed mass (top) and reconstruction loss (bottom) of the \model model on a simulated sample of QCD background (black) and multiple signals with different masses and decay chains (colored). No signal sample has been used in the training of the model. The mass of the intermediate particle is chosen to be half the mass of the parent particle.}
\label{fig:performance}
\end{figure}

The performance of the model is shown in Figure~\ref{fig:performance} for a variety of signals with different masses and decay structures. The model is able to correctly reconstruct all signals without bias regardless of the mass and decay, despite having been trained purely on background. The selection of $H_T > 1000$ GeV drives the background distribution to peak at around $H_T/2 \approx 500$ GeV. 
The reconstruction loss of the autoencoder block provides an excellent discriminating feature to suppress the background, achieving background rejection factors of $\mathcal{O}(200)$ with 50\% signal efficiency for the lowest mass points.
The loss is found to be strongly correlated with the reconstructed mass. Multiple possibilities exist to reduce or remove the correlation \cite{Louppe:2016ylz,Shimmin:2017mfk,Kitouni:2020xgb,Kasieczka:2020yyl,Klein:2022hdv,Algren:2023spv}, but are left for future work. 

\begin{figure}[t]
\centering
\includegraphics[width=0.45\textwidth]{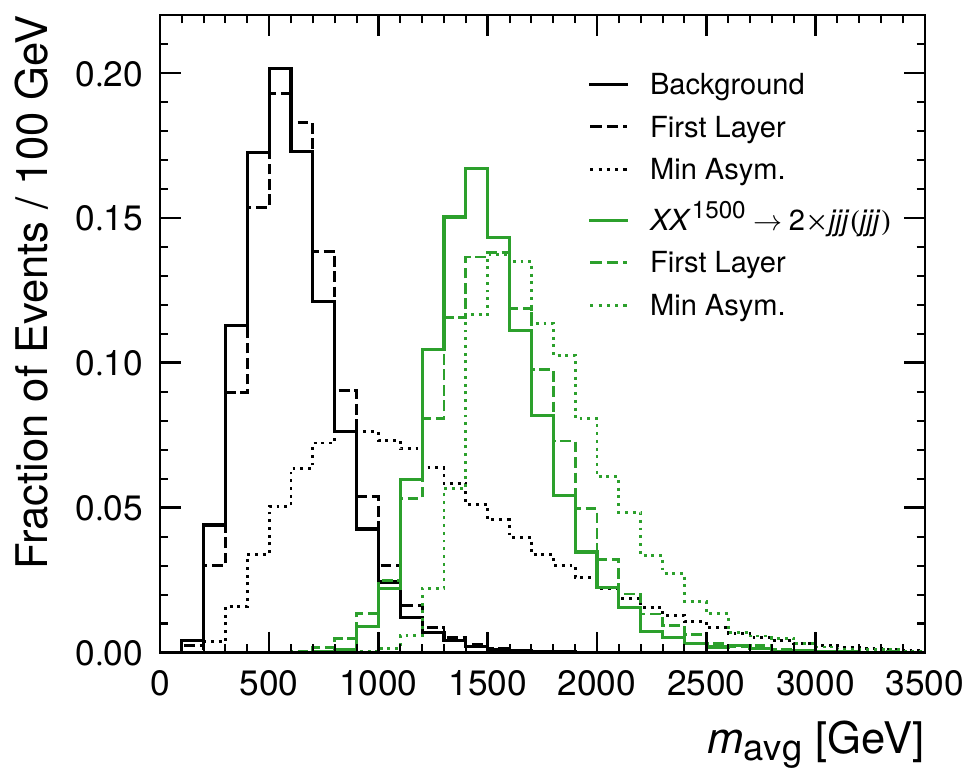} 
\caption{Comparison of the average mass for the background (black) and an example signal (green) as reconstructed by \model in its last layer (solid line), intermediate layer (dotted line), and the mass reconstructed from the minimization of the mass asymmetry (dashed line). An improvement of the signal resolution is observed after the first layer prediction is refined. The background distribution is strongly sculpted towards high masses when reconstructed by minimizing the mass asymmetry. }
\label{fig:compare}
\end{figure}

The performance of the model is further inspected using the jet assignment right after the first \abc layer to build the parent candidates. As shown in Figure~\ref{fig:compare}, the resolution of the reconstructed mass improves from the first to the second layer as the jet classification is refined. 
The performance is further compared with a heuristic approach such as iterating over all possible jet combinations and selecting the one that minimizes the mass asymmetry. The reconstructed signal mass exhibits worse resolution, a positive bias, and a large high-mass tail. The reconstructed background spectrum is shaped strongly towards high masses, leading to a factor 30 worse signal-to-background ratio under the signal peak, and a factor 200 worse when integrating to the highest mass. 

In conclusion, this letter presents a pioneering methodology that expands the landscape of possible BSM physics searches at the LHC. We develop the model-agnostic \model, based on \abc layers, which is suited for unsupervised learning directly on data and addresses critical limitations of current approaches. Our method not only obviates the need for predefined signal models but also demonstrates remarkable resilience in decoding complex, high-multiplicity final states. The ability to reconstruct pair-produced particles without knowledge of the decay mode is a significant leap in the pursuit of model-independent searches. 
The presented model, while falling in the category of anomaly-based searches, provides handles for the inspection of a possible signal through the explicit reconstruction from its decay products, which would allow the measurement of its mass, decay structure, and other properties. Looking forward, we encourage the experimental collaborations to leverage this method to expand their suite of model-independent searches into final states with high object multiplicity. The code is publicly available at \url{https://github.com/badeaa3/unsupervised-search}.



We thank Lawrence Lee, Simone Francescato, and Aurelio Juste Rozas for useful discussions and suggestions. This work has been supported by the Department of Energy, Office of Science, under Grant No. DE-SC0007881 (A.B.), the Harvard Graduate Prize Fellowship (A.B.), the Eric and Wendy Schmidt AI in Science Postdoctoral Fellowship (A.B.), grant RYC2021-030944-I funded by MCIN/AEI/10.13039/501100011033 (J.M.B) and by the European Union NextGenerationEU/PRTR (J.M.B).

\bibliography{Unsupervised} 

%
\end{document}